\documentclass[english]{achemso}
\usepackage[T1]{fontenc}
\usepackage[latin9]{inputenc}
\usepackage{geometry}
\usepackage{color}
\usepackage{amsmath}
\usepackage{amssymb}
\usepackage{graphicx
}

\usepackage{soul}

\newcommand*{\citen}[1]{%
  \begingroup
    \romannumeral-`\x 
    \setcitestyle{numbers}%
    \cite{#1}%
  \endgroup   
}
\makeatletter

\title{Optical Activity from the Exciton Aharonov-Bohm Effect: A Floquet Engineering
Approach}

\author{Kai Schwennicke\textsuperscript{1} and Joel Yuen-Zhou\textsuperscript{1}}
\email{joelyuen@ucsd.edu}

\affiliation{\textsuperscript{1}Department of Chemistry and Biochemistry, University
of California, San Diego, La Jolla, CA 92093, USA}

\makeatother

\usepackage{babel}
\begin{document}
\begin{abstract}
Floquet engineering is a convenient strategy to induce nonequilibrium
phenomena in molecular and solid-state systems, or to dramatically
alter the physicochemical properties of matter, bypassing costly and
time-consuming synthetic modifications. In this article, we investigate theoretically some interesting consequences of the fact that an originally achiral molecular system can exhibit
nonzero circular dichroism (CD) when it is driven with elliptically
polarized light. More specifically, we consider an isotropic ensemble
of small cyclic molecular aggregates in solution whose local low-frequency
vibrational modes are driven by an elliptically polarized continuous-wave infrared pump. We
 attribute the origin of a nonzero CD signal to time-reversal symmetry breaking
due to an excitonic Aharonov-Bohm (AB) phase arising from a time-varying laser electric field, together with coherent interchromophoric exciton hopping. The obtained Floquet engineered
excitonic AB phases are far more tunable than analogous magnetically-induced electronic
AB phases in nanoscale rings, highlighting a virtually unexplored potential
that Floquet engineered AB phases have in the coherent control of molecular
processes and simultaneously introducing new analogues of magneto-optical
effects in molecular systems which bypass the use of strong magnetic
fields.
\end{abstract}

\section{Introduction}

Molecular enantiomers exhibit slightly different interactions with
right and left-handed circular polarized (RCP, LCP) light fields.
There are two important consequences of this effect: non-racemic samples
exhibit finite circular dichroism (CD) (difference in absorption rate
of RCP and LCP fields) and finite optical rotation (OR) (rotation
of the plane of linearly polarized light). The phenomena of CD and
OR are intimately related by Kramers-Kronig relations and are collectively
known as ``optical activity'' \cite{barron2009molecular,lucarini2005kramers}. CD and
OR are well-established spectroscopic techniques to probe enantiomeric
excess in chemical samples. Interestingly, optical activity is not
restricted to samples involving chiral molecules, but can also arise
in a material that interacts with a static magnetic field parallel to the propagation direction of light. To distinguish
the optical activity due to a magnetic field from that due to the breakdown
of molecular inversion symmetry, the former so-called magneto-optical
(MO) effects are known as magnetic CD (MCD) and magnetic OR (MOR) 
or Faraday effect, respectively. MCD is an important spectroscopic
technique that allows one to resolve electronic transitions in congested
absorption bands\cite{thorne1977applications,jones1999freeze}, while MOR is a phenomenon
that is routinely exploited in the fabrication of optical devices
such as isolators and circulators\cite{thorne1977applications,ying2018strong,shoji2014magneto}.

\textcolor{black}{The fundamental origin of MO effects is the breakdown
of time-reversal symmetry (TRS). It is therefore reasonable that phenomena analogous to MO effects could arise from replacing a static magnetic field
with time-varying electric fields; we shall hereafter term them pseudo-MO
effects. In molecular spectroscopy, these phenomena have been long recognized
since the pioneering work of Atkins and Miller\cite{atkins1968quantum}.
More recently, within the context of topological photonics and nonreciprocal
media, much attention has been directed towards exploiting time-varying modulation
of material permittivities as a way to generate synthetic gauge (``pseudo-magnetic'')
fields, thereby circumventing magnetic field strengths that could
be prohibitive in general experimental setups or difficult to integrate
in optoelectronic devices. In fact, photonic versions of the Aharonov-Bohm
(AB) effect\cite{aharonov1959significance} and quantum Hall systems\cite{klitzing1980new,ando1975theory,novoselov2007room}
 have been experimentally demonstrated. The the AB effect was observed for photons using a Ramsey-type interferometer\cite{tzuang2014non}, while topologically protected photonic edge states were observed in a arrays constructed from optical ring resonators\cite{hafezi2013imaging}. A hallmark
of TRS breaking is the realization of AB phases that differ from 0
or $\pi$ $\text{mod}\,2\pi$. This version of AB phase is astutely realized in Reference   \citen{fang2012realizing}
by driving resonators in adjacent unit cells so that there is a large
phase lag between them, a feature that is facilitated by the relatively
big mesoscopic lengthscales of the experimental system of interest. Thus, it is
not obvious }\textcolor{black}{\emph{a priori}} 
whether such a strategy would be feasible in the recreation of pseudo-magnetic
fields in molecular aggregates or in general nanoscale scenarios,
especially considering the lengthscale mismatch between the typical
interatomic or intermolecular distances and the optical wavelength of the driving field. However, as we shall show in the present article, such a mismatch can
be compensated by invoking rings of coherently coupled anisotropic
nanoscale molecular dipoles which simultaneously interact with an elliptically polarized laser
field, such that different molecules interact with distinct linear polarizations
of the laser and consequently pick up distinct phases from the light to break TRS. Our work follows a similar spirit to other theoretical proposals where an optically induced AB effect can be achieved in both electronic\cite{sigurdsson2014optically} and excitonic\cite{kibis2015aharonov,kozin2018periodic}   nanoscopic ring structures by exploiting circularly polarized electric fields to break TRS. However, in these cases, the AB phase varies with field strength, while in our case the AB phase solely depends on the phase imprinted by the ellipticity of the field. 

In this paper, we present a theoretical proof-of-principle of the
possibility of inducing synthetic gauge fields in the excitonic degrees
of freedom of small cyclic molecular aggregates. In particular, we
design a setup consisting of an isotropic ensemble of
molecular homotetramers, where each of the sites has an internal vibronic
structure with low-frequency (e.g., vibrational) transitions that
are driven by a near-resonant electric field. Owing to the periodicity
of the laser-driving, we employ Floquet theory \cite{shirley1965solution,chu1989generalized,
tannor2007introduction}
to calculate the spectral properties of the driven system according
to a recently developed methodology\cite{gu2018optical}. These Floquet
engineering concepts are complementary to efforts using strong
magnetic fields to dramatically alter the vibronic structure and consequently
the optical properties of molecular aggregates\cite{maiuri2018high}, as well as to the design of topologically
nontrivial phases in organic excitonic systems\cite{yuen2014topologically}. Although by no means a new technique, Floquet engineering is regaining interest in the development of materials with novel properties\cite{oka2019floquet}, and in the proposal of using of Floquet engineering to influence both energy\cite{thanh2018control} and electron transfer\cite{phuc2019control} in molecular systems. In particular, our study highlights the simplicity with which large
values of AB phases can be Floquet engineered with time varying electric fields in nanoscale systems
(as shown below, with weak laser intensities), in contrast with the
difficulties of deploying giant magnetic fields to generate similarly large values of AB phases. Hence, our results suggest much flexibility in the use
of elliptically polarized light in the realization of excitonic AB
phases in molecular systems, which could control a variety of charge and energy transfer
processes in molecular aggregates or manipulate the spectroscopic
properties of the latter. While excitonic AB phases that depend on
the separation between electrons and holes have been studied in inorganic
Wannier exciton systems \cite{romer2000aharonov,
hu2001aharonov,govorov2002polarized,palmero2005aharonov,
kuskovsky2007optical},
similar phases in Frenkel organic exciton systems are far less explored (see Reference \citen{yuen2014topologically} for an example)
and constitute an interesting frontier in physical chemistry.

\section{Theoretical Formalism}

\begin{figure}
\includegraphics[scale=2]{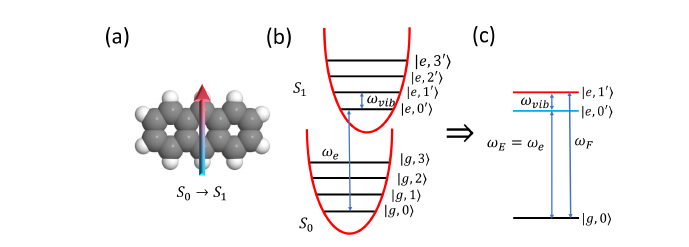}\caption{\emph{Anthracene monomer}. (a) Molecular structure and dipole moment
vector corresponding to the $S_{0}\to S_{1}$ transition. (b) Simplified
displaced harmonic oscillator spectrum of anthracene along low-frequency
vibrational mode ($\omega_{vib}=385\,\text{cm}^{-1}$). Here, $\omega_{e}=27,695\,\text{cm}^{-1}$
is the 0-0' electronic transition frequency and $D=0.31$ is the Huang-Rhys
parameter for vibronic coupling. (c) In this work, we simplify
the spectrum in (b) with a three-level model.}
\label{molecule}
\end{figure}

\begin{figure}
\includegraphics[]{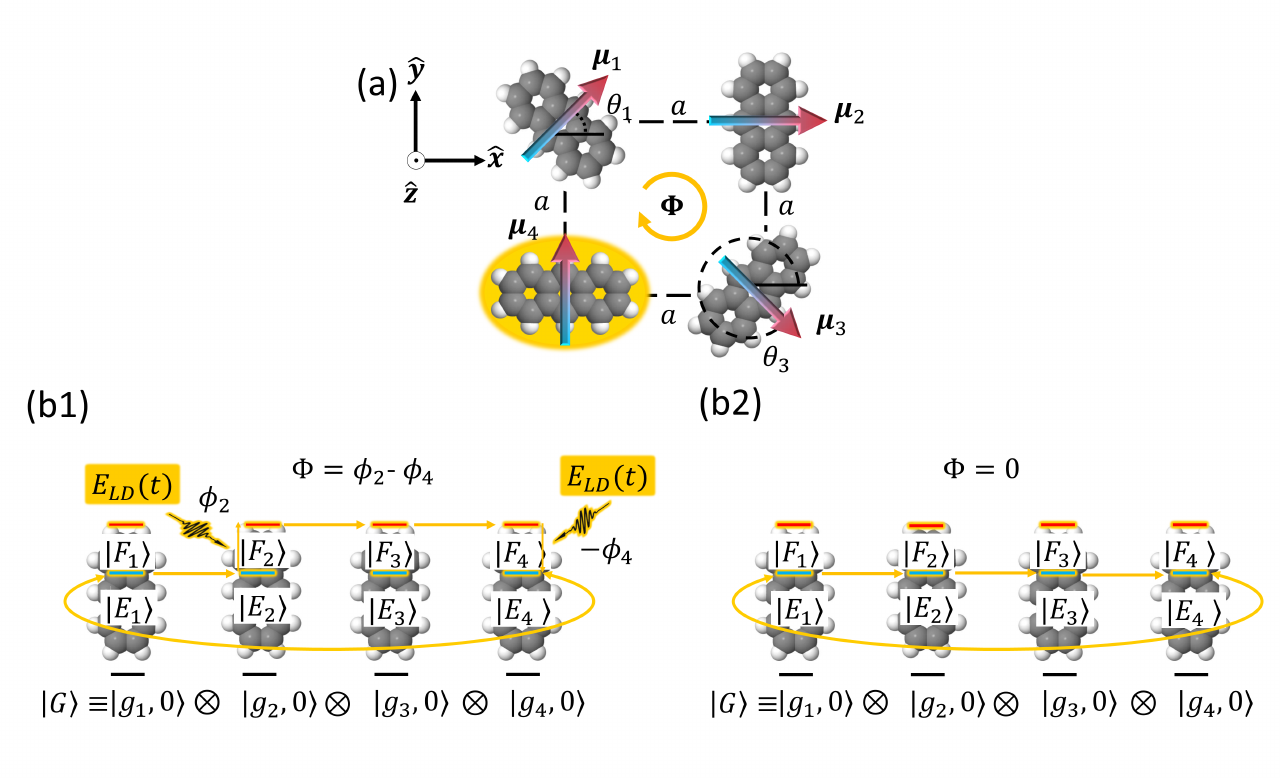}\caption{\emph{Realization of excitonic Aharonov-Bohm (AB) effect in cyclic
molecular aggregates via Floquet engineering with elliptically polarized
light.} (a) Example geometry of an anthracene homotetramer where relative
orientations ($\theta_{1}$, $\theta_{3}$) of transition dipoles
together with laser-driving induces an excitonic AB phase $\Phi$.
The anthracene monomers are equidistantly placed at a distance $a$ from their neighbors
forming a square. (b) Schematic illustration of the mechanism whereby
Floquet engineering induces excitonic AB phases. The elliptically
polarized field $E_{LD}(t)$ drives the $|E_{i}\rangle\to|F_{i}\rangle$
low-frequency vibrational transition (vertical yellow arrows). Horizontal
yellow arrows denote excitonic couplings mediated by electrostatic
interactions. In (b1), excitation of the first chromophore at $|E_{1}\rangle$
is resonantly transferred to the second chromophore $|E_{2}\rangle$
via a dipolar coupling. Next, the laser promotes the $|E_{2}\rangle\to|F_{2}\rangle$
vibrational excitation, ``dialing'' phase $\phi_{2}$ onto that
transition. Subsequent excitation transfer $|F_{2}\rangle\to|F_{3}\rangle\to|F_{4}\rangle$
occurs via resonant dipolar couplings, after which the laser promotes
the $|F_{4}\rangle\to|E_{4}\rangle$ vibrational emission, dialing
phase $-\phi_{4}$ onto that transition. Finally another step of resonant
dipole-mediated excitation transfer $|E_{4}\rangle\to|E_{1}\rangle$
closes the loop, yielding an AB phase $\Phi=\phi_{2}-\phi_{4}$.
$\Phi\protect\neq n\pi$ for integer $n$ signals time-reversal
symmetry breaking. Meanwhile (b2) illustrates a pathway mediated by electrostatic interactions (and no influence from laser-driving) that does not
result in nontrivial AB phase.}
\label{the_concept}

\end{figure}

\subsection{Definition of excitonic model}

The building block of our molecular aggregates is an anthracene molecule,
depicted in Figure \ref{molecule}. We shall only be concerned with
its $S_{0}\to S_{1}$ electronic transition (with 0-0' frequency $\omega_{e}=27,695\,\text{cm}^{-1}$)
which is coupled to a low-frequency $a_{g}$ mode ($\omega_{vib}=385\,\text{cm}^{-1}$);
the Huang-Rhys factor that characterizes the displacement of the $S_{1}$
potential energy surface with respect to the $S_{0}$ surface is $D=0.31$\cite{lambert1984jet}.
These parameters were used in previous theoretical studies addressing
how to control excitation-energy transfer by manipulating ultrafast
vibrational dynamics in anthracene dimers\cite{biggs2009calculations,biggs2012studies};
the values were fit to experimental data involving vibrational
coherences in anthracene dimers\cite{yamazaki2004observation}. The
weak vibronic coupling featured by this transition allows us to simplify
the anthracene spectrum as an effective three-level system featuring
ground $|g,0\rangle$ and excited $|e,0'\rangle$ electronic states
with no phonons, and an excited $|e,1'\rangle$ electronic state with
one phonon, where $|g\rangle$ and $|e\rangle$ denote $S_{0}$ and
$S_{1}$ electronic states, and $|\nu\rangle$ and $|\nu'\rangle$
label vibrational eigenstates of the harmonic potentials corresponding
to $S_{0}$ and $S_{1}$, respectively (see Figure \ref{molecule}).

For simplicity, let us consider a homotetramer where the chromophores
are located at the vertices of a square of side length $a=3.5\,\mathring{\text{A}}$,
mimicking values reported from the aforementioned ultrafast spectroscopy study\cite{yamazaki2004observation} (see Figure \ref{the_concept}). To unclutter notation, we introduce the following
single-excitation basis: \begin{subequations}\label{eq:exciton_basis}

\begin{align}
|G\rangle & =\prod_{i=1}^{4}|g_{i},0_{i}\rangle,\label{eq:G}\\
|E_{j}\rangle & =|e_{j},0'_{j}\rangle\prod_{i\neq j}^{4}|g_{i},0_{i}\rangle,\label{eq:Ej}\\
|F_{j}\rangle & =|e_{j},1'_{j}\rangle\prod_{i\neq j}^{4}|g_{i},0_{i}\rangle.\label{eq:Fi}
\end{align}
\end{subequations}\noindent The excitonic Hamiltonian of the homotetramer
reads ($\hbar=1$)

\begin{align}
H_{T}= & \sum_{i=1}^{4}\Big(\omega_{E}|E_{i}\rangle\langle E_{i}|+\omega_{F}|F_{i}\rangle\langle F_{i}|\Big)\nonumber \\
 & +\sum_{\langle ij\rangle}\Big[(J_{E_{i},E_{j}}|E_{i}\rangle\langle E_{j}|+J_{E_{i},F_{j}}|E_{i}\rangle\langle F_{j}|+J_{F_{i},F_{j}}|F_{i}\rangle\langle F_{j}|)+\text{h.c.}\Big],\label{eq:H_T}
\end{align}
where the sum over $\langle ij\rangle$ assumes only nearest-neighbor
couplings. Here $\omega_{E}=\omega_{e}$ and $\omega_{F}=\omega_{e}+\omega_{vib}=28,080\text{ cm}^{-1}$,
and $J_{\alpha\beta}=J_{\beta\alpha}$ are the electrostatic couplings
between excitonic states $|\alpha\rangle$ and $|\beta\rangle$, which
are calculated assuming the Condon approximation\cite{nitzan2006chemical,heller2018semiclassical}, e.g.

\[
J_{E_{i},F_{j}}=\langle e_{i}g_{j}|H_{T}|g_{i}e_{j}\rangle\langle0'_{i}|0_{i}\rangle\langle0_{j}|1'_{j}\rangle,
\]
where
\[
\langle e_{i}g_{j}|H_{T}|g_{i}e_{j}\rangle = \eta\frac{(\hat{\boldsymbol{\mu}}_{i}\cdot\hat{\boldsymbol{\mu}}_{j})-3(\hat{\boldsymbol{\mu}}_{i}\cdot\hat{\boldsymbol{r}}_{ij})(\hat{\boldsymbol{\mu}}_{j}\cdot\hat{\boldsymbol{r}}_{ij})]}{a^{3}}
\]
is approximated as the classical dipole-dipole interaction, with $\hat{\boldsymbol{\mu}_{i}}$
being the unit vector of the electronic transition dipole moment for the $S_{0}\to S_{1}$ transition
for the $i$-th chromophore and $\hat{\boldsymbol{r}}_{ij}$ being the unit vector that connects
chromophores $i$ and $j$. The value $\eta=982\,\text{cm}^{-1}\,\mathring{\text{A}}^{3}$
has been invoked to account for the effects of transition dipole magnitudes as well as the index of refraction of the
surrounding medium, so that the dipolar coupling between parallel
anthracene molecules is $\langle e_{i}g_{j}|H_{T}|g_{i}e_{j}\rangle=\frac{\eta}{a^{3}}=22.9\,\text{cm}^{-1}$\cite{biggs2009calculations,biggs2012studies},
a value that is consistent with quantum beat data in ultrafast spectroscopy
experiments\cite{yamazaki2004observation}.
The scalar $|\langle n'_{i}|0_{i}\rangle|^{2}=\text{exp}[-D]\frac{D^{n}}{n!}$
is the relevant Franck-Condon factor\cite{nitzan2006chemical}. In
this work, we fix the orientations of $\boldsymbol{\mu}_{2}$ and
$\boldsymbol{\mu}_{4}$ to be along the $\hat{\boldsymbol{x}}$ and
$\hat{\boldsymbol{y}}$ axes, respectively, in the molecular aggregate frame, while
we vary the orientations of $\boldsymbol{\mu}_{1}$ and $\boldsymbol{\mu}_{3}$
at angles $\theta_{1}$ and $\theta_{3}$ (Figure \ref{the_concept}a
shows $\theta_{1}=45^{0}$ and $\theta_{3}=315^{0}$). For simplicity,
we ignore out-of-plane dipole orientations; this is not a necessary
condition and could be easily lifted if chemical synthesis favors
other geometries. The excitonic transition dipole moments from the ground state, $|G\rangle$, are $\boldsymbol{\mu}_{E_{i}G} =\langle E_{i}|\boldsymbol{\mu}|G\rangle=\langle e_{i},0'_{i}|\boldsymbol{\mu}|g_{i},0_{i}\rangle$ and $\boldsymbol{\mu}_{F_{i}G}=\langle F_{i}|\boldsymbol{\mu}|G\rangle=\langle e_{i},1'_{i}|\boldsymbol{\mu}|g_{i},0_{i}\rangle$. In our calculations we take $\langle e_{i},0'_{i}|\boldsymbol{\mu}|g_{i},0_{i}\rangle=0.90\,\text{Debye}$ and $\langle e_{i},1'_{i}|\boldsymbol{\mu}|g_{i},0_{i}\rangle=0.74\,\text{Debye}$, values which are consistent with experiments\cite{yamazaki2004observation}.

\begin{figure}
\includegraphics[scale=0.5]{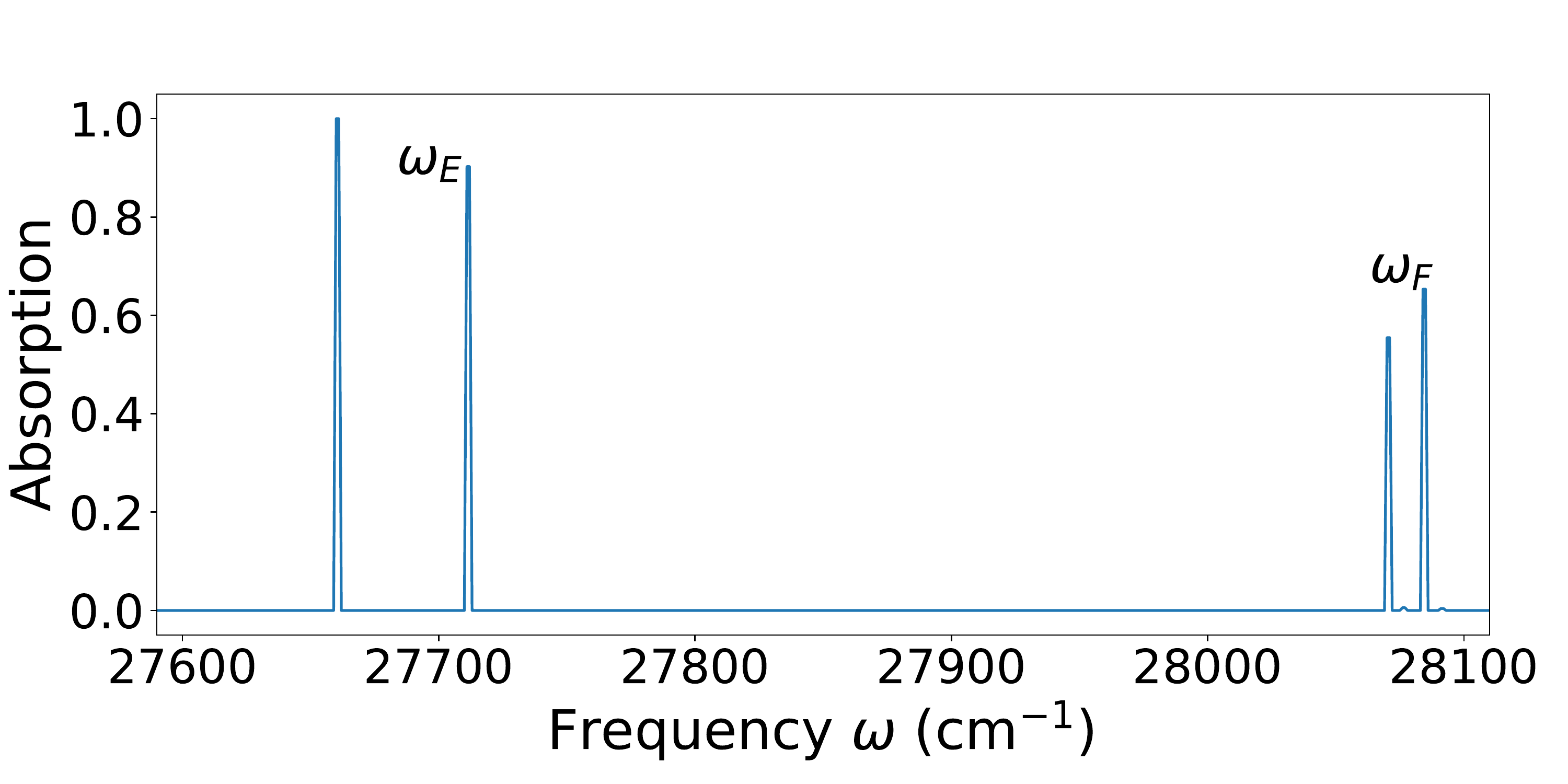}\caption{\emph{Absorption spectrum of an isotropic solution of excitonic homotetramers.}
It can be approximately understood as consisting of two bands of
transitions corresponding to coherent combinations of 0-0' transitions
(centered at $\omega_{E}$) and 0-1' transitions (centered at $\omega_{F}$) respectively.
The splitting of each of these bands is due to resonant exciton hopping between
molecules, and correspond to $\sim2\text{max}|J_{E_{i}E_{j}}|$ and
$\sim2\text{max}|J_{F_{i}F_{j}}|$, respectively.}

\label{no_pump_absorbance}
\end{figure}

Given that $J_{\alpha\beta}\ll\omega_{vib}$, the exciton eigenstates
of $H_{T}$ consist of two bands centered at $\omega_{E}$ and $\omega_{F}$,
respectively, with bandwidths $\sim2\text{max}|J_{E_{i}E_{j}}|,\,\sim2\text{max}|J_{F_{i}F_{j}}|$.
Figure \ref{no_pump_absorbance} shows the absorption spectrum of
an isotropically averaged collection of homotetramers. Notice that
due to different Franck-Condon overlaps, the splitting of the band at $\omega_{F}$
is narrower than that at $\omega_{E}$. Note, there are multiple vibronic eigenstates whose symmetries result in net zero transition dipole moments; that is, they are dark with respect to UV-visible light in the absence of laser-driving. Accordingly, these states cannot be observed using absorption spectroscopy, which is why there are fewer peaks in the spectrum than the number  of possible eigenstates of $H_{T}$. 

\subsection{Floquet theory for laser-driving}

The aforementioned two bands can be mixed upon introduction of a driving
laser; the new Hamiltonian is time-dependent and reads

\begin{equation}
H_{LD}(t)=H_{T}-\sum_{i=1}^{4}(\boldsymbol{\mu}_{E_{i}F_{i}}\cdot\boldsymbol{E}_{LD}(t)|E_{i}\rangle\langle F_{i}|+\text{h.c.}),\label{eq:H_LD}
\end{equation}
where we have introduced an infrared (IR) field at a frequency $\Omega=\omega_{vib}+\delta$
that is slightly detuned from the $E_{i}\to F_{i}$ vibrational transitions
by $\delta$:

\[
\boldsymbol{E}_{LD}(t)=\frac{E_{LD}^{0}}{\sqrt{2}}[\hat{x}\cos(\Omega t+\phi_{x})+\hat{y}\cos(\Omega t+\phi_{y})].
\]
The detuning $\delta$ allows us to ignore resonant pumping of vibrational
transitions in the ground state (Figure \ref{molecule}b) so that
the three-level approximation per molecule (Figure \ref{molecule}c)
is justified. The phases $\phi_{x}$ and $\phi_{y}$ will play an
important role below. The relevant transition dipole moments to the
laser-driving are $\boldsymbol{\mu}_{F_{i}E_{i}}=\langle F_{i}|\boldsymbol{\mu}|E_{i}\rangle=\langle e_{i},1'_{i}|\boldsymbol{\mu}|e_{i},0'_{i}\rangle$
and we take $|\boldsymbol{\mu}_{F_{i}E_{i}}|=0.15\,\text{Debye}$,
which is a reasonable value for IR molecular vibrational excitations\cite{maj2015beta}.

For completeness, we shall now lay out the Floquet theory\cite{shirley1965solution,chu1989generalized,
tannor2007introduction} 
utilized to address the problem in question. The Hamiltonian $H_{LD}(t)$
is periodic in time with period $T=2\pi/\Omega$ such that $H_{LD}(t+T)=H_{LD}(t).$
The solutions to the time dependent Schr{\"o}dinger equation (TDSE)
\[
i\frac{d}{dt}|\Psi(t)\rangle=H_{LD}(t)|\Psi(t)\rangle
\]

\noindent can be written in the Floquet state basis $|\Psi(t)\rangle=\sum_{\lambda}C_{\lambda}|\psi_{\lambda}(t)\rangle$
where
\begin{equation}
|\psi_{\lambda}(t)\rangle=e^{-i\varepsilon_{\lambda}t}|\phi_{\lambda}(t)\rangle.\label{eq:floquet_states}
\end{equation}

\noindent The Floquet modes can be spectrally decomposed in the single
(molecular) excitation basis, $\alpha\in\{G,E_{i},F_{i}\}$ where $i\in\{1,2,3,4\}$, introduced in Eq. (\ref{eq:exciton_basis}),

\noindent
\begin{equation}
|\phi_{\lambda}(t)\rangle=\underset{\alpha}{\sum}|\alpha\rangle\langle\alpha|\phi_{\lambda}(t)\rangle ,\label{eq:floquet_modes}
\end{equation}
and are periodic in time with period $T$. Furthermore, they are eigenfunctions
of the Floquet Hamiltonian
\begin{equation}
H_{F}|\phi_{\lambda}(t)\rangle=\Bigg[H_{LD}(t)-i\frac{d}{dt}\Bigg]|\phi_{\lambda}(t)\rangle=\varepsilon_{\lambda}|\phi_{\lambda}(t)\rangle,\label{eq:Hf}
\end{equation}
where the eigenvalues $\{\varepsilon_{\lambda}\}$ are known as quasi-energies.
Since $|\phi_{\lambda}(t)\rangle$ and $H_{LD}(t)$ are time-periodic,
we can write Eqs. (\ref{eq:H_LD}) and (\ref{eq:floquet_modes})  in
terms of their Fourier components:

\noindent \begin{subequations}\label{eq:Fourier_basis}
\begin{align}
|\phi_{\lambda}(t)\rangle & =\underset{\alpha}{\sum}\sum_{n=-\infty}^{\infty}|\alpha\rangle\langle\alpha|\phi_{\lambda}^{(n)}\rangle e^{in\Omega t},\label{eq:fm_fb}\\
\langle\alpha|H_{LD}(t)|\beta\rangle & =\sum_{n=-\infty}^{\infty}H_{LD,\alpha\beta}^{(n)}e^{in\Omega t},\label{eq:hf_fb}
\end{align}
\end{subequations}\noindent where $H_{LD,\alpha\beta}^{(n)}=\frac{1}{T}\int_{0}^{T}dt\langle\alpha|H_{LD}(t)|\beta\rangle e^{-in\Omega t}$.
By substituting Eqs. (\ref{eq:fm_fb}) and (\ref{eq:hf_fb}) into
Eq. (\ref{eq:Hf}), multiplying both sides of the resulting equation
by $\exp(-ik\Omega t)$, and taking a time-integral over one period
$T$, we obtain

\noindent \begin{subequations}\label{eq:Shirley}
\begin{align}
\sum_{\beta n}H_{F\alpha\beta}^{(k-n)}\langle\beta|\phi_{\lambda}^{(n)}\rangle & =\varepsilon_{\lambda}\langle\alpha|\phi_{\lambda}^{(k)}\rangle,\label{eq:new_eig_problem}\\
H_{F\alpha\beta}^{(k-n)} & =H_{LD,\alpha\beta}^{(k-n)}+n\Omega\delta_{\alpha\beta}\delta_{kn}.\label{eq:transformed_HF}
\end{align}

\noindent \end{subequations}\noindent At this point, we augment the
Hilbert space from $|\alpha\rangle$ to $|\alpha\rangle\otimes|t\rangle$
, where $\langle t|\phi_{\lambda}\rangle=|\phi_{\lambda}(t)\rangle$,
and introduce the Fourier basis $\{|n\rangle\}$ where $\langle n|\phi_{\lambda}\rangle\equiv|\phi_{\lambda}^{(n)}\rangle$,
$\langle t|n\rangle=e^{in\Omega t}$, $|\alpha,n\rangle\equiv|\alpha\rangle|n\rangle$, and
$\langle\alpha,k|H_{F}|\beta,n\rangle\equiv H_{F\alpha\beta}^{(k-n)}$, rendering the eigenvalue problem into the following form:

\noindent \begin{subequations}\label{eq:Shirley_2}

\begin{align}
H_{F}|\phi_{\lambda}\rangle & =\Big[H_{F}^{(0)}+H_{F}^{(1)}\Big]|\phi_{\lambda}\rangle=\varepsilon_{\lambda}|\phi_{\lambda}\rangle,\label{eq:Hf_fin_a}\\
H_{F}^{(0)}  = &\sum_{n=-\infty}^{\infty}\Big\{\sum_{i=1}^{4}\Big[(\omega_{E}+n\Omega)|E_{i},n\rangle\langle E_{i},n|+(\omega_{F}+n\Omega)|F_{i},n\rangle\langle F_{i},n|\Big]+n\Omega|G,n\rangle\langle G,n|\nonumber \\
 & +\sum_{\langle ij\rangle}\Big[(J_{E_{i},E_{j}}|E_{i},n\rangle\langle E_{j},n|+J_{E_{i},F_{j}}|E_{i},n\rangle\langle F_{j},n|+J_{F_{i},F_{j}}|F_{i},n\rangle\langle F_{j},n|)+\text{h.c.}\Big]\Big\},\label{eq:Hf_fin_b}\\
H_{F}^{(1)}  = &-\frac{E_{LD}^{0}}{\sqrt{2}}\sum_{n=-\infty}^{\infty}\Big\{\sum_{i=1}^{4}\sum_{q=\hat{\boldsymbol{x}},\hat{\boldsymbol{y}}}\Big[\mu_{E_{i}F_{i}}^{q}\big(e^{i\phi_{q}}|E_{i},n\rangle\langle F_{i},n+1|+e^{-i\phi_{q}}|E_{i},n+1\rangle\langle F_{i},n|\big)+\text{h.c}\Big]\Big\}.\label{eq:Hf1}
\end{align}
\end{subequations}\noindent An intuitive interpretation of this basis,
which was originally presented by Shirley\cite{shirley1965solution},
comes from associating $n$ to the number of quanta present in the
laser-driving mode; e.g., $|E_{1},3\rangle$ would refer to the state
were the exciton corresponds to the 0-0' transition in first chromophore
while the field contains three photons. Following this interpretation,
Eq. (\ref{eq:Hf_fin_b}) denotes that intermolecular interactions between excitonic
states preserve photon number, while Eq. (\ref{eq:Hf1}) shows
that the driving field couples states $|E_{i}\rangle,|F_{i}\rangle$
by either absorbing or emitting a photon. To gain further insight,
we consider the limit where
\begin{equation}
|J_{E_{i}F_{j}}|,|\boldsymbol{\mu}_{LD}\cdot\boldsymbol{E}_{LD}(t)|\ll\Omega,\label{eq:RWA_condition}
\end{equation}
so that we can apply the rotating-wave approximation (RWA) to drop
highly off-resonant terms; Eqs. (\ref{eq:Hf_fin_b}) and (\ref{eq:Hf1})
can then be reduced to

\begin{subequations}\label{eq:RWA}
\begin{align}
H_{F}^{(0)} & \approx\sum_{n=-\infty}^{\infty}\Big\{\sum_{i=1}^{4}\Big[(\omega_{E}+n\Omega)|E_{i},n\rangle\langle E_{i},n|+(\omega_{F}+n\Omega)|F_{i},n\rangle\langle F_{i},n|\Big]+n\Omega|G,n\rangle\langle G,n|\nonumber\\
 & +\sum_{\langle ij\rangle}\Big[(J_{E_{i},E_{j}}|E_{i},n\rangle\langle E_{j},n|+J_{F_{i},F_{j}}|F_{i},n\rangle\langle F_{j},n|)+\text{h.c.}\Big]\Big\}\label{eq:Hf_0_RWA} \\
H_{F}^{(1)} & \approx-\frac{E_{LD}^{0}}{\sqrt{2}}\sum_{n=-\infty}^{\infty}\Big\{\sum_{i=1,q}^{4}\Big[\mu_{E_{i}F_{i}}^{q}e^{-i\phi_{q}}|E_{i},n+1\rangle\langle F_{i},n|
+\text{h.c.}\Big]\Big\}.\label{eq:Hf_1_RWA}
\end{align}
\end{subequations}\noindent The RWA invokes a physically intuitive
constraint: the laser-driving intensity is weak enough that it can
only induce the $|E_{i}\rangle\to|F_{i}\rangle$ transition if the
exciton absorbs a photon from the field. Furthermore, the RWA block-diagonalizes
Eq. (\ref{eq:Hf_fin_a}) so that $H_{F}\approx\underset{n}{\sum}( h_{F,n}+h_{G,n})$
where the $n$-th blocks $h_{G,n}$ and $h_{F,n}$ are defined as:
\noindent \begin{subequations}\label{eq:block}

\begin{align}
h_{G,n} & = n\Omega |G,n\rangle \langle G,n|,\label{eq:h_Gn}\\
h_{F,n} & =\sum_{i=1}^{4}\Big[(\omega_{E}+(n+1)\Omega)|E_{i},n+1\rangle\langle E_{i},n+1|+(\omega_{F}+n\Omega)|F_{i},n\rangle\langle F_{i},n|\Big]\nonumber \\
 & +\sum_{\langle ij\rangle}\Big[(J_{E_{i},E_{j}}|E_{i},n+1\rangle\langle E_{j},n+1|+J_{F_{i},F_{j}}|F_{i},n\rangle\langle F_{j},n|)+\text{h.c.}\Big]\nonumber \\
 & -\frac{E_{LD}^{0}}{\sqrt{2}}\sum_{i=1,q}^{4}\Big[\mu_{E_{i}F_{i}}^{q}e^{-i\phi_{q}}|E_{i},n+1\rangle\langle F_{i},n|+\text{h.c}\Big].\label{eq:h_Fn}
\end{align}
\end{subequations}\noindent Note that the Floquet states $\{|\psi_{\lambda}(t)\rangle\}$ are
uniquely characterized by the quasi-energies $\varepsilon_{\lambda}\,\mod\Omega$\cite{tannor2007introduction,shirley1965solution}; therefore, within the RWA, we only need to diagonalize one $h_{F,n}$ block and one $h_{G,n}$ block 
to find a finite set of Floquet modes $\{|\phi_\lambda \rangle\}$ to construct the Floquet state basis $\{|\psi_{\lambda}(t)\rangle\}$. We  
should note that Eq. (\ref{eq:h_Fn}) resembles a lattice Hamiltonian
where the quantum degrees of freedom are influenced by a gauge field
$\vec{\boldsymbol{A}}(\boldsymbol{r})$ \cite{hofstadter1976energy},
where $\phi_{q}=\int\vec{\boldsymbol{A}}(\boldsymbol{r})\cdot d\boldsymbol{r}$.
Typically, the latter arises from a magnetic field coupling to movable charges, resulting in AB
phases; in our case, we shall see that the gauge field is a result
of the elliptically polarized laser, which can be thought of as a pseudo-magnetic field.

\subsection{Calculation of circular dichroism spectrum}

To calculate the CD signal of the laser-driven system, we invoke the formalism
developed in Reference \citen{gu2018optical}. We compute the rates of absorption
$W_{\pm}$ due to weak-intensity continuous-wave circularly polarized
probe laser fields at UV-visible frequency $\omega$,

\[
\boldsymbol{E}_{P\pm}(\omega,t)=\frac{E_{P}^{0}}{\sqrt{2}}(\hat{\boldsymbol{x}}\cos\omega t\pm\hat{\boldsymbol{y}}\text{sin}\omega t).
\]
The CD response at each frequency $\omega$ is defined as the difference
$\delta W(\omega)=W_{+}(\omega)-W_{-}(\omega)$ in the rates due to RCP and LCP light,

\noindent
\begin{equation}
\delta W(\omega)=-|E_{p}^{0}|^{2}\pi\text{Im}\Bigg[\sum_{\lambda,\alpha,\beta}\sum_{n=-\infty}^{\infty}\mu_{\alpha G}^{y}\langle\phi_{\lambda}|\alpha n\rangle\mu_{G\beta}^{x}\langle\beta n|\phi_{\lambda}\rangle\delta(\varepsilon_{\lambda}-n\Omega-\omega)\Bigg].\label{eq:cd_full}
\end{equation}

\begin{figure}
\includegraphics[]{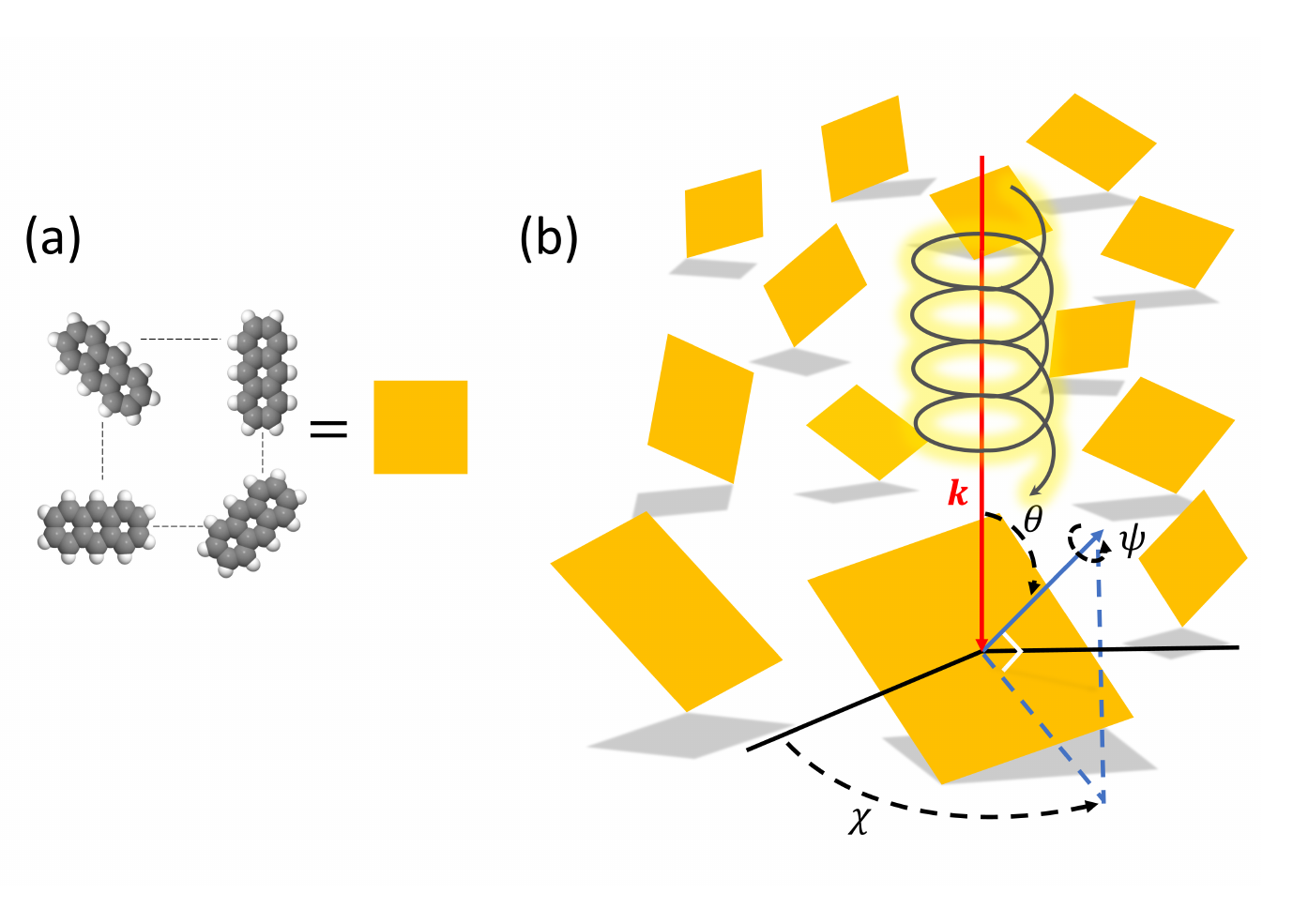}

\caption{(a) Excitonic homotetramer unit. (b) Depiction of isotropic ensemble
of homotetramers; calculation of circular dichroism (CD) involves
averaging over orientations $\chi,\theta,\psi$ with respect to the
$\boldsymbol{k}$ vector of the incident beam. Here, the normal vector to the molecular aggregate plane (blue arrow) is uniquely defined by the polar coordinates $\theta$ and $\chi$, while $\psi$ is the angle of rotation of the aggregate about the aforementioned vector. Note that the ellipticity of the incident beam is not represented to scale with respect to the molecular aggregates.}

\label{orientations}
\end{figure}

\noindent Here $\varepsilon_{\lambda}-n\Omega$ denotes the $|Gn\rangle\to|\phi_{\lambda}\rangle$ transition energy, where $|\phi_{\lambda}\rangle$ is an eigenstate
of $h_{F,1}$ or $h_{G,1}$ (see Eqs. (\ref{eq:h_Gn}), (\ref{eq:h_Fn})) with eigenvalue $\varepsilon_{\lambda}$, and
$\omega$ is the energy absorbed from the probe. Note that $\boldsymbol{\mu}_{GG}=0$ and, due to
the RWA, $h_{F,1}$ only couples the states $|E_{i},2\rangle$ and $|F_{i},1\rangle$. Therefore, $\langle\alpha,n|\phi_{\lambda}\rangle\neq0$ only for $n=1,2$, which results in
Eq. (\ref{eq:cd_full}) being simplified to
\begin{align}
\delta W(\omega) & =-|E_{p}^{0}|^{2}\pi\text{Im}\Bigg[\sum_{\lambda}\sum_{ij}\Big\{\mu_{E_{i}G}^{y}\langle\phi_{\lambda}|E_{i}2\rangle\mu_{GE_{j}}^{x}\langle E_{j}2|\phi_{\lambda}\rangle\delta(\varepsilon_{\lambda}-2\Omega-\omega)\label{eq:Cd_RWA}\\
 & +\mu_{F_{i}G}^{y}\langle\phi_{\lambda}|F_{i}1\rangle\mu_{GF_{j}}^{x}\langle F_{j}1|\phi_{\lambda}\rangle\delta(\varepsilon_{\lambda}-\Omega-\omega)\Big\}\Bigg].\nonumber
\end{align}
 Eq. (\ref{eq:Cd_RWA}) corresponds to the CD for a single homotetramer of fixed orientation. Since we are interested in an isotropic film
or solution of such aggregates, we must take an average of $\delta W(\omega)$
over different orientations of the tetramer with respect to the incident
light $\boldsymbol{k}$ vector,

\noindent
\begin{equation}
\langle\delta W(\omega)\rangle=\frac{1}{8\pi^{2}}\int_{0}^{2\pi}\int_{0}^{2\pi}\int_{0}^{\pi}d\chi d\psi d\theta W(\omega;\chi,\psi,\theta)\text{sin}\theta,\label{eq:isotropic}
\end{equation}

\noindent where we have explicitly written $\delta W(\omega)=\delta W(\omega;\chi,\psi,\theta)$
to express the fact that the CD for each aggregate is a function of
its orientation, defined by Tait-Bryan angles $\chi ,\,\psi ,\,\theta$\cite{baranowski2013equations} (see Figure \ref{orientations}). Notice that unlike with perturbative
spectroscopy\cite{mukamel1995principles}, we cannot analytically
carry out this average given that $h_{F,1}$ is a function itself
of $\chi ,\,\psi , \, \theta$, as each orientation experiences a different
driving due to its different dipole projections with the IR laser.
Hence, we compute the isotropically averaged CD spectrum in Eq. (\ref{eq:isotropic})
via Monte Carlo integration (see Supplementary Information).

\section{{\large{}Results and Discussion}}

\subsection{Excitonic AB phases}

To illustrate the nontrivial effects produced by Floquet engineering, we
now consider coherent pathways due to the various terms in Eq. (\ref{eq:block}).
In particular, we are interested in cyclic ones, namely, those
which begin and end in the same state. For concreteness, let us focus
on the pathway depicted in Figure \ref{the_concept}b1. Excitation
of the first chromophore at $|E_{1}\rangle$ is resonantly transferred
to the second chromophore $|E_{2}\rangle$ via dipolar coupling. Next,
the elliptically polarized driving laser promotes the $|E_{2}\rangle\to|F_{2}\rangle$
vibrational excitation, imprinting a nontrivial phase $\phi_{2}$
onto that transition. Subsequent excitation transfer via dipolar coupling
$|F_{2}\rangle\to|F_{3}\rangle\to|F_{4}\rangle$ is followed by a
laser-induced vibrational de-excitation $|F_{4}\rangle\to|E_{4}\rangle$,
which imprints yet another nontrivial phase $\phi_{4}$; the pathway
is closed by another dipolar coupling $|E_{4}\rangle\to|E_{1}\rangle$.
To make these statements more precise, let us define the Wilson loop\cite{wilson1974confinement}
corresponding to this pathway:

\begin{align*}
\mathcal{W} & =\langle E_{1}n|h_{F,n}|E_{4}n+1\rangle\langle E_{4}n+1|h_{F,n}|F_{4}n\rangle\langle F_{4}n|h_{F,n}|F_{3}n\rangle\\
& \times\langle F_{3}n|h_{F,n}|F_{2}n\rangle\langle F_{2}n|h_{F,n}|E_{2}n+1\rangle\langle E_{2}n+1|h_{F,n}|E_{1}n+1 \rangle\\
& =J_{E_{1}E_{2}}\Bigg(\sum_{q=\hat{\boldsymbol{x}},\hat{\boldsymbol{y}}}\frac{E_{LD}^{0}\mu_{E_{2}F_{2}}^{q}}{\sqrt{2}}e^{i\phi_{q}}\Bigg)J_{F_{2}F_{3}}\\
& \times J_{F_{3}F_{4}}\Bigg(\sum_{q=\hat{\boldsymbol{x}},\hat{\boldsymbol{y}}}\frac{E_{LD}^{0}\mu_{E_{2}F_{2}}^{q}}{\sqrt{2}}e^{-i\phi_{q}}\Bigg)J_{E_{4}E_{1}}.
\end{align*}
If we define

\begin{align*}
\phi_{2}&=\text{arg}\text{\ensuremath{\Bigg[J_{E_{1}E_{2}}\Bigg(\sum_{q=\hat{\boldsymbol{x}},\hat{\boldsymbol{y}}}\frac{E_{LD}^{0}\mu_{E_{2}F_{2}}^{q}}{\sqrt{2}}e^{i\phi_{q}}\Bigg)J_{F_{2}F_{3}}\Bigg]}},\\
\phi_{4}&=-\text{\ensuremath{\arg\Bigg[J_{F_{3}F_{4}}\Bigg(\sum_{q=\hat{\boldsymbol{x}},\hat{\boldsymbol{y}}}\frac{E_{LD}^{0}\mu_{E_{2}F_{2}}^{q}}{\sqrt{2}}e^{-i\phi_{q}}\Bigg)J_{E_{4}E_{1}}\Bigg]}},
\end{align*}
the excitonic AB phase corresponding to this pathway is equal to $\Phi=\arg(W)=\phi_{2}-\phi_{4}$.
To gain further intuition on the types of transition dipole arrangements
that lead to substantial values of $\Phi$, let us consider the configuration
in Figure \ref{the_concept}a, where the dipoles in chromophores 2
and 4 are aligned along $\hat{\boldsymbol{x}}$ and $\hat{\boldsymbol{y}}$ respectively,
and the only positive dipolar couplings within our gauge convention
is $J_{F_{2}F_{3}}$; then, $\phi_{2}=\phi_{x}+\pi$ and $\phi_{4}=-\phi_{y}-2\pi=-\phi_{y}$,
so that the phases of the elliptically polarized field are explicitly
imprinted in the resulting excitonic AB phase
\[
\Phi=\arg(\mathcal{W})=\phi_{x}-\phi_{y}+\pi=\Delta\phi+\pi.
\]
A few comments are pertinent at this point. When $\Delta\phi=0$~$\text{mod\,\ensuremath{\pi}},$ the
driving laser corresponds to linearly polarized light and the AB phase
$\Phi=0,\pi$ is trivial; this is expected as TRS is preserved. However,
when $\Delta\phi\neq0$~$\text{mod\,\ensuremath{\pi}}$, the driving
laser corresponds to elliptically polarized light, leading to a nontrivial
AB phase $\Phi\neq0,\pi$ ($\Delta\phi=\frac{\pi}{2}$~$\text{mod\,\ensuremath{\pi}}$
corresponds to the special case of circular polarization) that signals
TRS breaking. Hence, our design crucially depends on chromophores
2 and 4 obtaining different phases $\phi_{x}$ and $\phi_{y}$ from
the laser-driving; to maximize this difference, they are placed perpendicular
to one another. However, the production of a nontrivial AB phase relies
also on having coherent dipolar couplings throughout the aggregate
that yield a nonzero Wilson loop. To ensure that chromophores 2 and
4 are coupled, chromophores 1 and 3 are positioned along optimal orientations
$\frac{\hat{\boldsymbol{x}}\pm\hat{\boldsymbol{y}}}{\sqrt{2}}$ so
that the exciton hoppings are enough to ensure large coherent couplings
throughout the cycle. It follows that to realize an exciton AB phase
$\Phi=\frac{\pi}{2}$, one simply needs to set $\Delta\phi=-\text{\ensuremath{\frac{\pi}{2}}},$ corresponding
to right circularly polarized light. At this point, it is worth highlighting
that this excitonic AB phase is independent of laser intensity so
long as the RWA is a good approximation. This should be contrasted
with the magnitudes of the magnetic fields used to generate an AB phase $\Phi=\frac{\pi}{2}$ in a nanoscale ring supporting electronic currents.
In fact, considering a putative loop of the same area $a^{2}$ as
our molecular aggregate, $\Phi=\frac{eBa^{2}}{\hbar}$ \cite{aharonov1959significance}
would require $B=8400\,\text{Tesla}$ (an intensity that is only feasible
at present in restricted places such as white dwarfs!\cite{stopkowicz2015coupled}). This remark implies that it should be substantially
easier to realize AB phases in nanoscale systems by employing Floquet engineering via time-varying laser electric fields over giant static magnetic fields.
Thus, these versatile Floquet engineered AB phases could open doors to new ways of coherently controlling excitonic
processes in molecules and nanomaterials.

\textcolor{black}{Our proposal to induce AB phases via Floquet engineering
is heavily influenced by Reference \citen{fang2012realizing}, where the authors
suggest the periodic modulation of permittivities in arrays of coherently
coupled resonators to generate pseudo-magnetic fields. Crucial to their
setup is a well-crafted spatial distribution of modulation phases which is locally controlled
with electrical circuit elements. This local control of phases is
possible in mesoscopic systems but much harder to realize in the nanoscale.
In fact, one could be tempted to think that light-matter interaction
is ineffective in doing so, given the mismatch between the lengthscale
of light and molecules. However, as we show in our example above,
we achieve the distribution of modulation phases by exploiting (a)
the phase lag between the $\hat{\boldsymbol{x}}$ and $\hat{\boldsymbol{y}}$
components of the electric field of elliptically polarized laser light
and (b) by designing rings of coherently coupled molecular dipoles
where different sites feature orientations that interact with different
polarizations of the laser light.}

\subsection{Features of the CD spectrum}

\begin{figure}
\includegraphics[scale=0.45]{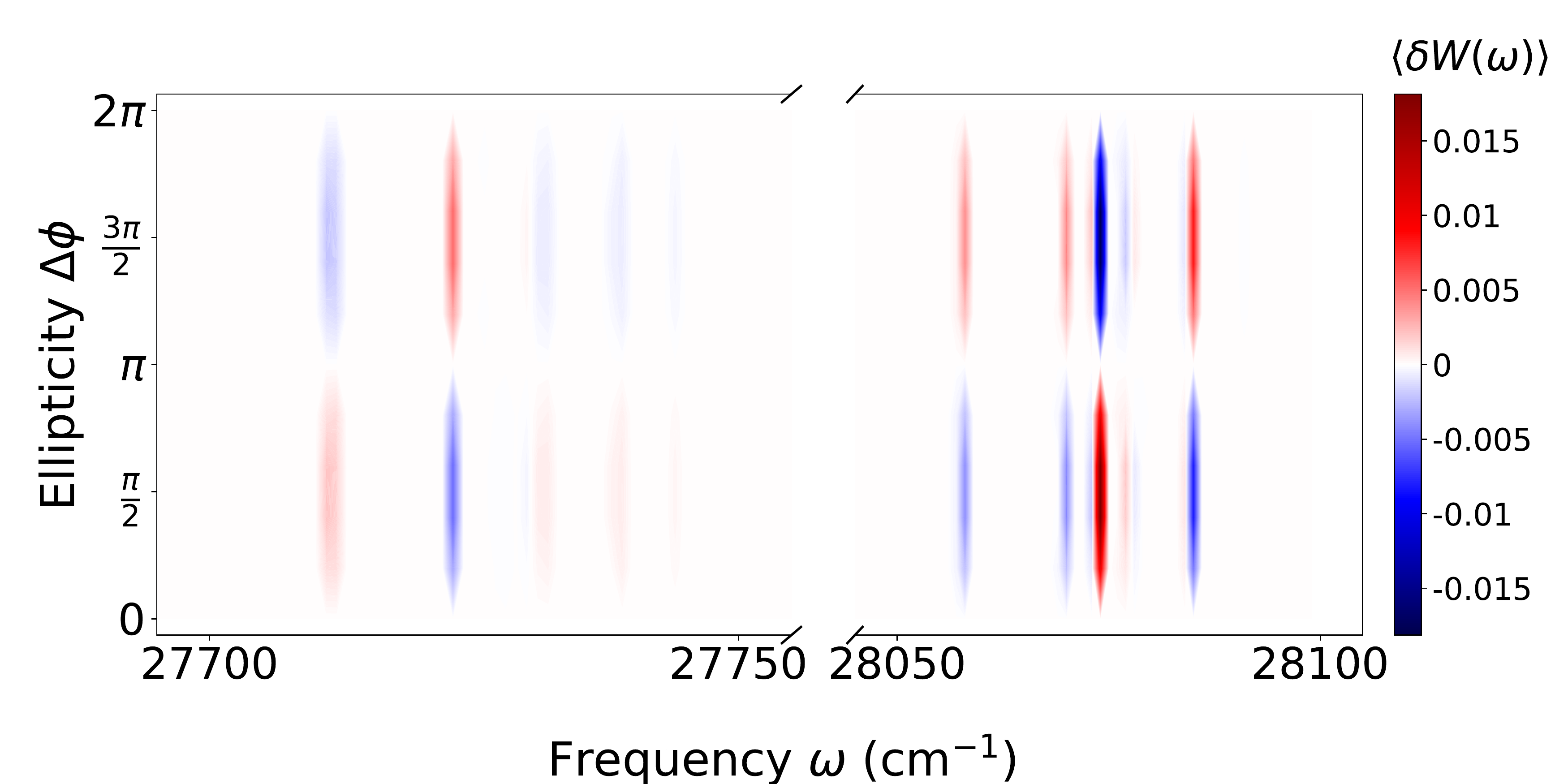}

\caption{Computed circular dichroism (CD) spectrum for isotropic ensemble of
molecular homotetramers under the influence of elliptically-polarized
driving, where the CD signal $\langle\delta W(\omega)\rangle$ is normalized with respect to the maximum absorption of the isotropic system when there
is no laser-driving. Clear progressions of the CD as a function of
ellipticity $\Delta\phi$ can be observed, with sign switches at the
time-reversal-symmetry (TRS) points $\Delta\phi=0,\pi$. Note we have inserted a break in the frequency axes in order to highlight the CD signal about $\omega_{E}$ and $\omega_{F}$.}

\label{CD}
\end{figure}

\noindent To test the effects of TRS breaking due to nontrivial excitonic
AB phases, Figure \ref{CD} shows the isotropic averaged CD spectrum
$\langle\delta W(\omega)\rangle$ normalized to the maximum absorption
of the isotropic solution in the absence of driving. For our simulations,
we use $E_{LD}^{0}=2.7\times10^{8}\,\frac{V}{m}$, which is a standard
value utilized in Floquet engineering experiments with inorganic semiconductors\cite{wang2013observation},
and take $\delta=-0.1\omega_{vib}$. With these parameters we obtain $\frac{|\boldsymbol{\mu}_{LD}\cdot\boldsymbol{E}_{LD}(t)|}{\Omega}=0.02$ ,
which satisfies the requirement of Eq. (\ref{eq:RWA_condition}) above.
As expected, when $\Delta\phi=0,\pi$, $\langle\delta W(\omega)\rangle$
vanishes for all $\omega$: TRS is preserved under linearly
polarized light driving. However, arbitrary elliptically polarized
fields ($\Delta\phi\neq0,\pi$) give rise to nonzero CD, where a clear
pattern of sign switches occurs at the TRS points. Notice that despite
the laser-driving being weak in comparison to its carrier frequency,
$|\boldsymbol{\mu}_{LD}\cdot\boldsymbol{E}_{LD}(t)|\ll\Omega$ (see  Eq. (\ref{eq:RWA_condition})), the
small shifts in peak frequencies in Figure \ref{no_pump_absorbance}
give rise to substantial CD signals, just as in MCD, where moderate
values of magnetic fields can easily dissect congested spectra\cite{thorne1977applications,jones1999freeze}. Also note that
the maximum absolute value of $ \langle\delta W(\omega)\rangle $ occurs
at $\Delta\phi=\frac{\pi}{2},\frac{3\pi}{2}$, which corresponds to
circularly polarized light. Based on the mechanism outlined in the
previous subsection (see also Figures \ref{the_concept}a, b), we
attribute this symmetry to the fact that in our homotetramer, the
dipoles are all of the same magnitude.

\section{{\large{}Conclusion}}

In this article, we have demonstrated that Floquet engineering with
elliptically-polarized laser fields can serve as an alternative to
magnetic fields to induce pseudo-MO effects in molecular systems and
nanomaterials. These phenomena should be regarded as complementary
to MCD and MOR effects in molecular systems. Small cyclic molecular
aggregates with anisotropic arrangement of transition dipoles can
support a wide range of excitonic AB phases upon elliptically polarized
laser-driving despite the small areas they enclose. A rich ellipticity-dependent
modulation of energy level splittings ensues, which concomitantly
manifests in widely tunable CD spectra.
Within the program of harnessing
coherence in light-harvesting systems\cite{scholes2017using, calderon2019nonadiabatic,tiwari2013electronic,chenu2013enhancement} and AB effects in quantum tunneling\cite{noguchi2014aharonov}, our
study emphasizes the potential that AB phases have in the coherent
control of energy and charge transport in molecular aggregates, as
well as in the decongestion of spectra of the latter. While we have
previously proposed the realization of excitonic AB phases in topologically protected
porphyrin arrays\cite{yuen2014topologically}, such effects have been
barely explored in a broader range of systems and they could be appealing
as a way to switch exciton couplings and propagation direction on demand with lasers rather than by synthetic modification. Moreover,
we have recognized that in the nanoscale, Floquet engineering is a dramatically less
challenging tool than using magnetic fields to  realize AB effects. In fact,
the nanoscale excitonic AB phases that arise in our designed system
depend on the pump laser ellipticity $\Delta\phi$ and not on its
field strength; thus a large phase of $\Phi=\frac{\pi}{2}$ can in
principle be readily realized in our protocol. This ease must be contrasted
with the prohibitive magnetic fields that must be used in nanorings
to induce an electronic AB phase of the same magnitude (we must mention,
however, ingenious molecular electronics proposals to harness small
AB phases arising from weak magnetic fields to induce substantial
control on electronic currents \cite{hod2006magnetoresistance}).
Furthermore, it is important to emphasize that while general features
of pseudo-MO and MO effects are similar (they both arise from breaking
of TRS), they cannot be easily compared\cite{atkins1968quantum}:
the former depends on coupling of time-varying electric fields with
electric transition dipole moments, while the latter arises from the
interaction of static magnetic fields with the spin and orbital angular
momentum of molecular eigenstates. Yet, we suspect that pseudo-MO
effects could provide a convenient alternative in situations where
magnetic fields are experimentally unfeasible.

The present article has laid the foundations of the generation of exciton AB phases in molecular aggregates. However, a detailed exploration of dissipative effects of a condensed phase environment must be addressed to ensure the experimental feasibility of our predictions in a wide range of experimental scenarios. Broadly speaking, we expect exciton AB phases to be resilient to decoherence as long as the light-matter coupling $|\boldsymbol{\mu}_{LD}\cdot\boldsymbol{E}_{LD}(t)|$ and the excitonic couplings $|J_{\alpha \beta}|$ are stronger than the spectral linewidths caused by the environment. These issues, together with the additional effects of intramolecular vibrations, will be addressed in future work.

\textit{Note.--} After initial submission of the manuscript, we were made aware of an important reference by Phuc and Ishizaki\cite{phuc2019control}, which proposes Floquet engineering to control the chirality of electron transfer in molecular trimers. While the conclusions in that work are similar to the present, the AB phase in their work is due to linearly polarized light with two driving frequencies and amplitudes, rather than due to elliptically polarized light. While no isotropic averaging effects were discussed in that work, the authors carried out a thorough exploration of the effects of decoherence and concluded that chiral currents survive in the condensed phase. Further comparison of both strategies to induce AB phases will be the subject of future studies.

\section{Supplementary Information Available}
Description of the Monte Carlo integration method used to calculate the CD for the driven homotetramer; presentation of the convergence of the CD signal calculation (Figure S1). This material is available free of charge via the Internet at http://pubs.acs.org.
\begin{acknowledgement}
K.S. and J.Y.Z. acknowledge support from the Air Force Office of Scientific
Research award FA9550-18-1-0289 for the excitonic design of AB phases
and the Defense Advanced Research Projects Agency under Award No.
D19AC00011 for the calculation of pseudo-MO effects. K.S. acknowledges
discussions with R.F. Ribeiro, M. Du, and S. Pannir-sivajothi. Both authors acknowledge incisive and helpful comments by Reviewer 1.
\end{acknowledgement}
\bibliography{IFE_bib}
\end{document}